\documentclass[11pt]{article}
\usepackage{moriond,epsfig,amsmath,amssymb}

\usepackage{xspace}

\newcommand{\mixing}{$D^0\!-\!\overline{D}^0$\xspace}

\newcommand{\rdcs}{R_{\mbox{\scriptsize{\textsc{dcs}}}}}

\newcommand{\mMev}{${\mathrm{MeV}}\!/c^2$\xspace}
\newcommand{\rarr}{\!\rightarrow\!}
\newcommand{\rs}{$D^0\rarr K^-\pi^+$\xspace}
\newcommand{\ws}{$D^0\rarr K^+\pi^-$\xspace}
\newcommand{\dzbar}{$\overline{D}^0$\xspace}
\newcommand{\dz}{$D^0$\xspace}
\newcommand{\cp}{$CP$\xspace}

\newcommand{\mix}{\mathrm{mix}}

\def\Journal#1#2#3#4{{#1} {\bf #2}, #3 (#4)}

\def\NIMA{{\em Nucl. Instrum. Methods} A}
\def\NPB{{\em Nucl. Phys.} B}
\def\PLB{{\em Phys. Lett.}  B}
\def\PRL{\em Phys. Rev. Lett.}
\def\PRD{{\em Phys. Rev.} D}

\begin{document}
\vspace*{4cm}
\title{$D^0-\bar{D}^0$ Mixing in FOCUS}

\author{Jonathan M. Link\\for the FOCUS Collaboration}

\address{Department of Physics, University of California, Davis, CA 95616, USA}

\maketitle
\abstracts{We report on a direct measurement of the mixing parameter
$y=(3.42\pm1.39\pm0.74)\%$ in the \mixing system by measuring the lifetime
difference between the \cp mixed final state $K^+\pi^-$ and the \cp even state
$K^+K^-$.  We also present a study of the decay \ws based on a sample of 
$149\!\pm\!31$ observed events compared to $36\,760\!\pm\!195$ events observed 
in the Cabibbo favored channel \rs. The observed branching ratio 
$R=(0.404\pm 0.085\pm 0.025)\%$ is used to obtain limits on the mixing
parameters $x^{\prime}$ and $y^{\prime}$ and the  doubly Cabibbo suppressed 
branching ratio, $\rdcs$.  These studies are based on a large sample of 
photoproduced charm mesons from the FOCUS experiment at Fermilab (FNAL-E831).}

\section{Introduction}
Mixing occurs because the neutral $D$ mass eigenstates (or \cp eigenstates in
the limit of \cp conservation) do not coincide with the flavor eigenstates \dz
and \dzbar.  The mixing effects are parameterized by two dimensionless 
amplitudes 
\begin{eqnarray}
x=\frac{\Delta M}{\Gamma} & \mbox{and} & y=\frac{\Delta \Gamma}{2\Gamma}, 
\end{eqnarray}
where $\Delta M$ is the mass difference between the two mass eigenstates, 
$\Delta\Gamma$ is the width difference and $\Gamma$ is the average width.  In 
the Standard Model, the \mixing system mixing rate
($R_{\mix}=\frac{1}{2}(x^2+y^2)$) is generally believed to be much smaller than 
the current experimental sensitivity~\cite{Nelson:1999fg}.  Nevertheless, 
recent measurements hint at a possible mixing effect at the edge of 
sensitivity.  We report here on two such studies made with the FOCUS data.

The data were collected by the FOCUS Collaboration during the 1996-97 Fermilab 
fixed target run in the Wideband Photon beam line using an upgraded version of 
the E687 spectrometer~\cite{Frabetti:1992au}.  Charm particles are produced in 
the interaction of high energy photons ($\langle E \rangle\!\approx\! 180$~GeV) 
with a segmented BeO target.  In the target region, charged particles are 
tracked by 16 layers of silicon microstrip detectors which provide excellent 
vertex resolution.  The momentum of the charged particles is determined by 
measuring their deflection in two oppositely polarized, large aperture dipole 
magnets with five stations of multiwire proportional chambers.  Particle 
identification is determined by three multicell threshold \v{C}erenkov 
detectors, electromagnetic calorimeters, and muon counters.

\section{Direct Measurement of $\Delta \Gamma$ from Lifetime Differences}
\label{dgamma}
\begin{figure}
\centerline{
\psfig{file=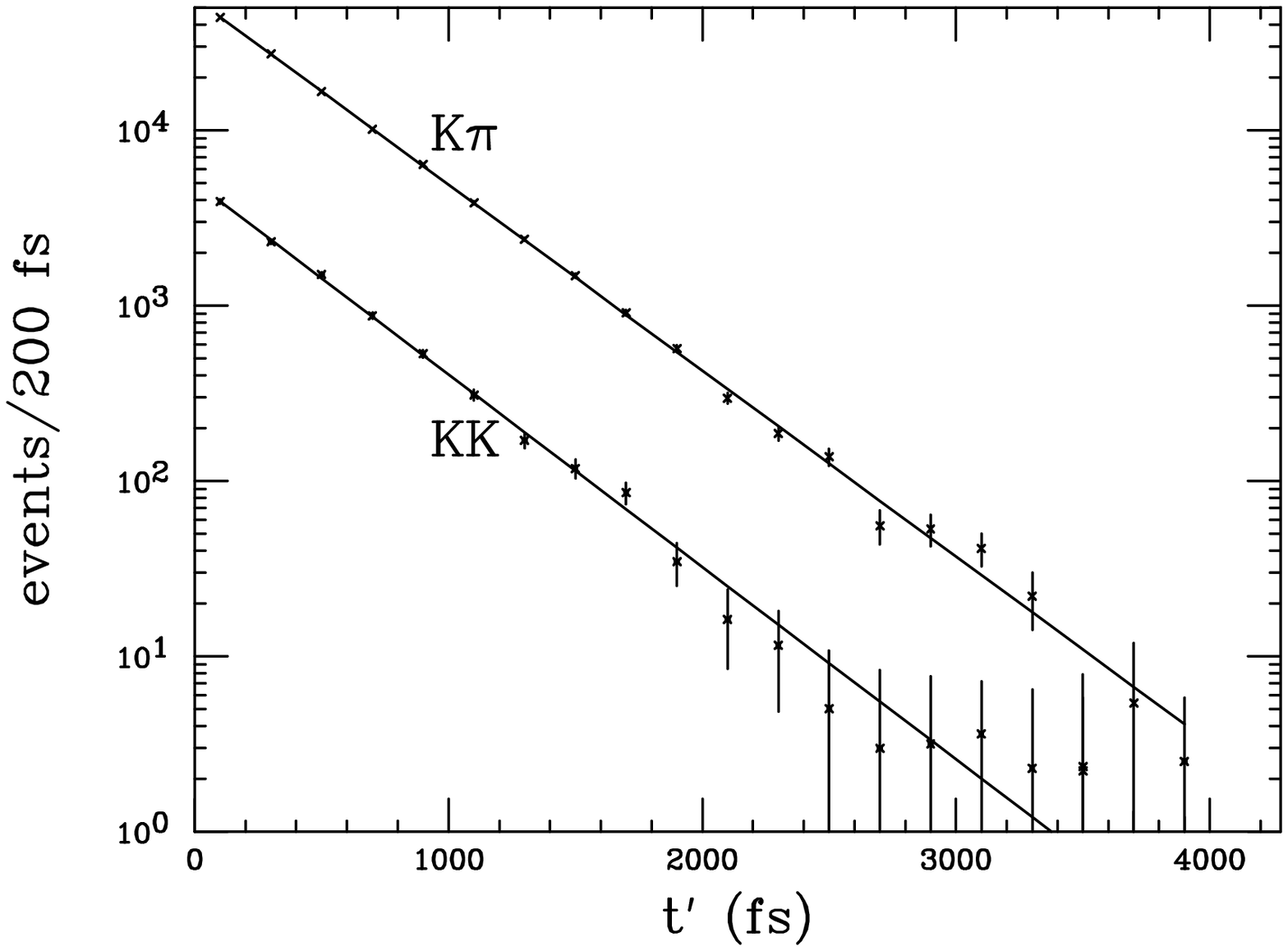,width=3.1in,height=2.8in}\hspace{0.3in}
\psfig{file=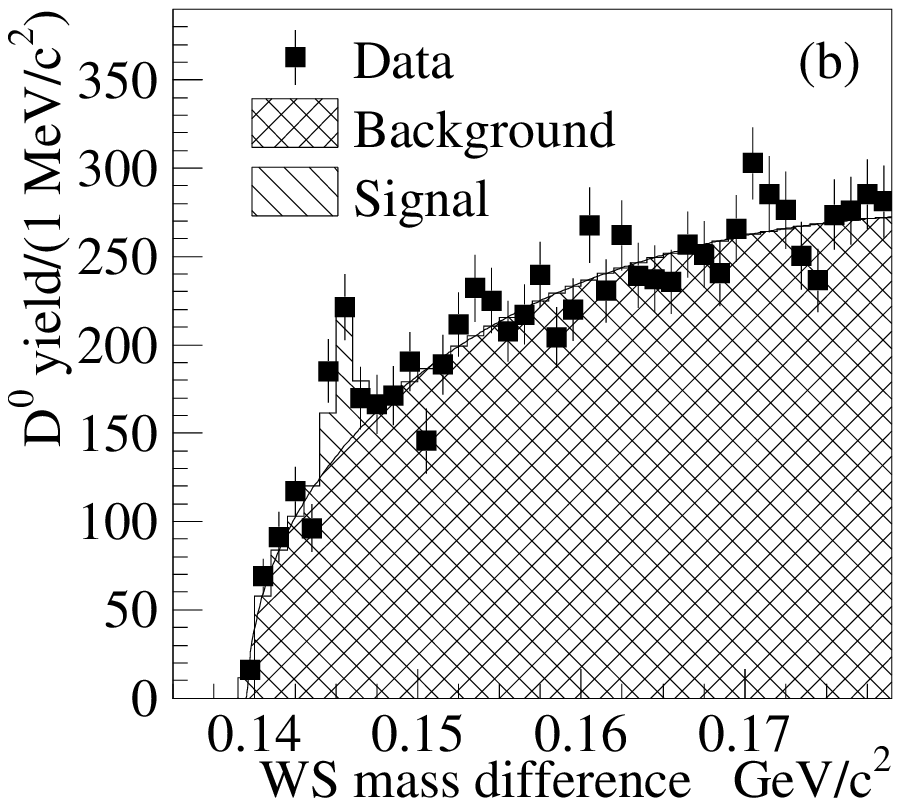,width=3.1in}}
\caption{(a) The decay reduced proper time distribution for the selected \rs 
and $K^+K^-$ events.  The data is background subtracted and includes a small 
Monte Carlo correction.  (b) The \ws mass difference distribution with the 
signal and background fit contributions shown.} 
\label{data}
\begin{picture}(0,0)
\put (205,240) {(a)}
\end{picture}
\end{figure}

By measuring and comparing the lifetime for neutral $D$'s decaying to final 
states of pure even and odd \cp a direct measurement of $\Delta \Gamma$ can be 
made.  In this study~\cite{Link:2000cu}, the final state $K^+K^-$ is used as 
the \cp even final state and, in the absence of a suitable \cp odd candidate, 
the \cp mixed state $K^-\pi^+$ is used.  Assuming that $K^-\pi^+$ is an even 
mixture of \cp even and \cp odd the relationship between the two lifetimes and 
the mixing parameter $y$ is given by
\begin{equation}
y=\frac{\Gamma_{CP\ even}-\Gamma_{CP\ odd}}{\Gamma_{CP\ even}+\Gamma_{CP\ odd}} 
= \frac{\tau(D\rarr K\pi)}{\tau(D\rarr KK)} -1.
\label{y}
\end{equation}
The lifetime distributions for the 10\,331 decays to $K^+K^-$ and
119\,738 decays to $K^-\pi^+$ are shown in Fig.~\ref{data}a.  From exponential 
fits to the distributions we find $\tau(D\rarr KK) = 395.7\pm5.5\ fs$ and 
$\tau(D\rarr K\pi) = 409.2\pm1.3\ fs$ where the errors are statistical
(systematic errors are only evaluated on the ratio).  Plugging these lifetimes 
into Eq.~\eqref{y} we obtain $y=(3.42\pm1.39\pm0.74)\%$.

\section{Study of the Decay \ws}
The process \ws may occur through either a doubly Cabibbo suppressed (DCS) 
decay, or by \dz mixing to \dzbar followed by the Cabibbo favored (CF) decay
$\overline{D}^0\rarr K^+\pi^-$.  The expected rate for the DCS decay relative
to the CF decay ($\rdcs$) is approximately $\tan^4\theta_{\mathrm{C}}\simeq
0.25\%$, while Standard Model based predictions for $R_{\mix}$ range from 
$10^{-8}$ to $10^{-3}$.  The large uncertainty in the \mixing mixing rate is due
to mixing mediated by intermediate meson states~\cite{Wolfenstein:1985ft} whose 
contributions are not calculable in perturbative QCD\@.  Nevertheless, large 
cancellations among the various intermediate meson states are expected and most 
studies conclude that the mixing rate should be at least a couple orders of 
magnitude less than $10^{-3}$ level~\cite{Donoghue:1986hh}.  Also, effects 
from beyond the Standard Model may enhance $R_{\mix}$.  

In this study~\cite{Link:2001kr}, we begin by measuring the rate of \ws decays
relative to \rs~\footnote{Charge conjugate modes are implicitly included.}.  The
neutral $D$ flavor is determined by requiring the decay chain $D^{+*}\rarr
\pi^+D^0\rarr \pi^+(K\pi)$.  This is achieved by looking for a narrow signal at
145~\mMev in the mass difference between the $D^*$ and $D$ candidates.  The mass
difference distribution for \ws candidates is shown in Fig.~\ref{data}b.  We 
find $149\pm 31$ \ws events compared to $36\,760\pm195$ decays of \rs for a
branching ratio of ($0.404\pm0.085\pm0.025$)\%.

\begin{figure}
\centerline{
\psfig{file=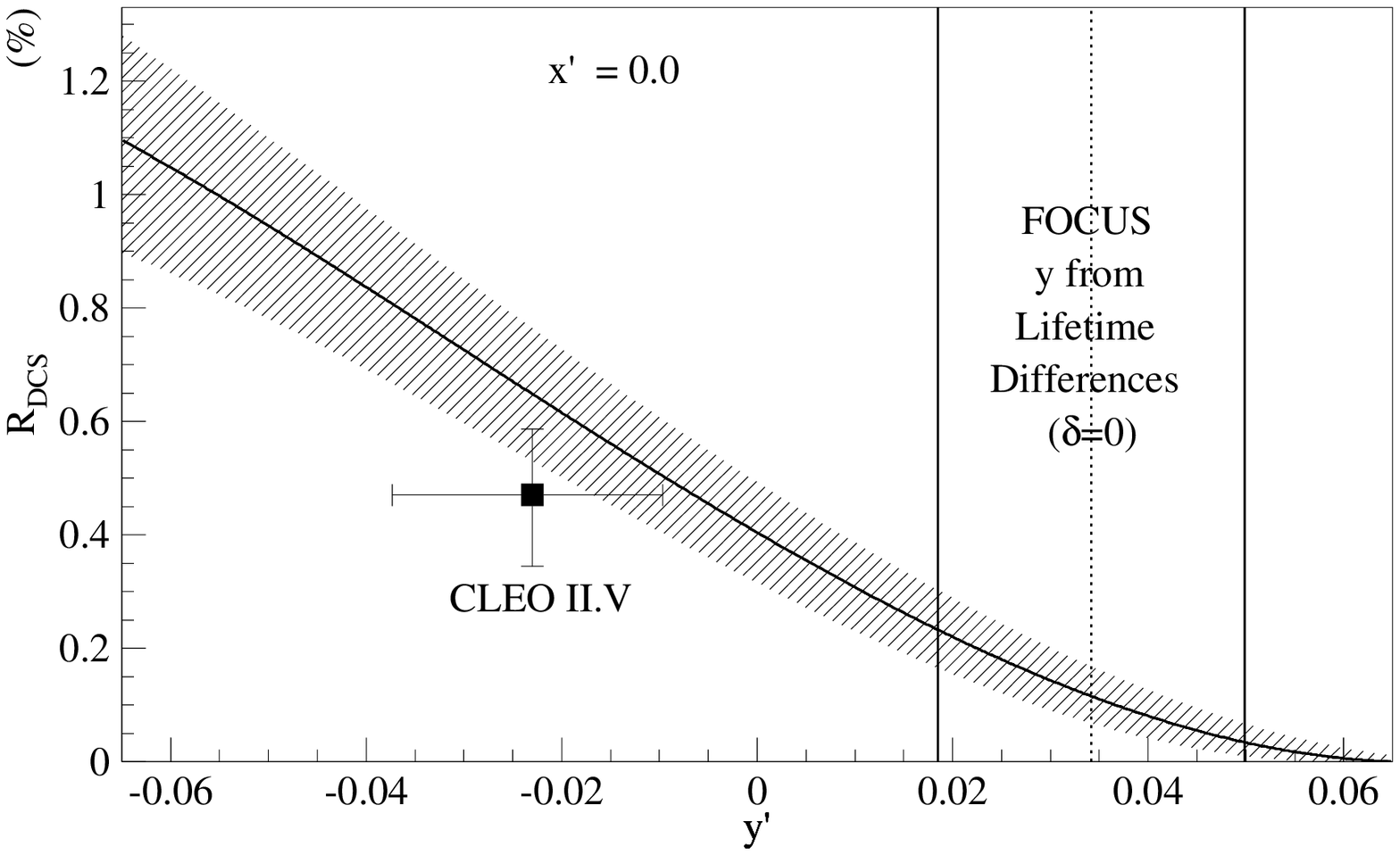,height=2.7in} \hspace{0.1in}
\psfig{file=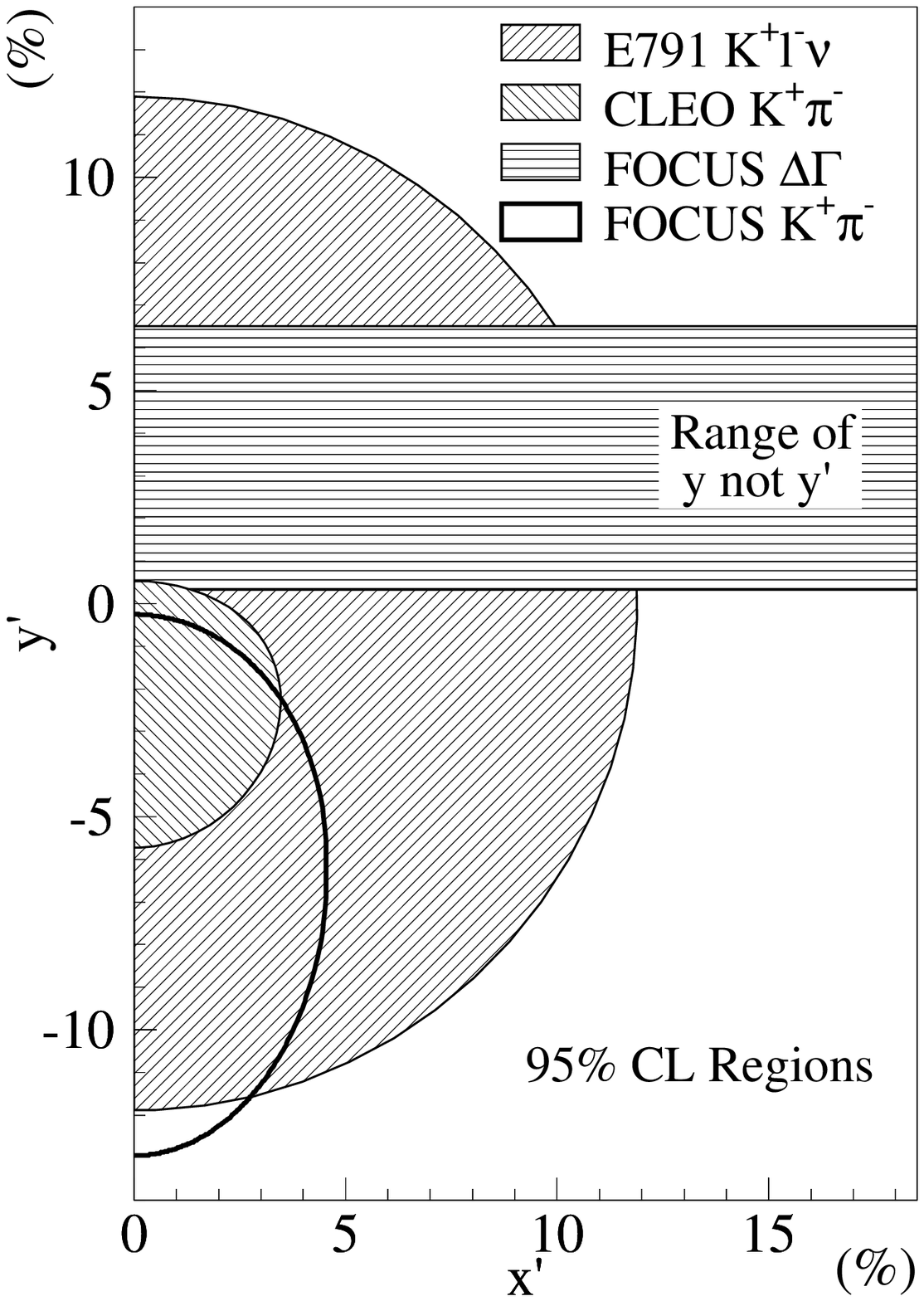,height=2.7in}}
\begin{picture}(0,0)
\put (50,195) {(a)}
\put (375,197) {(b)}
\put (438,165) {\scriptsize{Preliminary}}
\end{picture}
\caption{(a) The functional dependence of $\rdcs$ on $y^{\prime}$ with
$x^{\prime}=0$.  Fit values of $y^{\prime}$ and $\rdcs$ from 
CLEO~II.V~\cite{Godang:1999yd} and the the FOCUS value for $y$ from the 
lifetime difference study are shown for comparison.  All errors are $1\,\sigma$ 
combined statistical and systematic.  (b) The 95\% confidence level allowed 
regions for $x^{\prime}$ and $y^{\prime}$ from the E791 semileptonic 
study~\cite{Aitala:1996vz}, the CLEO~II.V study of \ws and the FOCUS study of 
\ws, and the FOCUS limits on $y$ from lifetime differences.  The direct 
measurement of $y$ is only comparable to measurements on $y^{\prime}$ in the 
limit of strong phase $\delta=0$.  For the measurements of $y^{\prime}$ and 
$y$ to be in agreement at the $1\,\sigma$ level requires $\delta \gtrsim 
\pi/4$~\cite{Bergmann:2000id}.}
\label{mixing}
\end{figure}

The time dependence of the \ws decays is given by
\begin{equation}
R(t) = \left[\rdcs + \sqrt{\rdcs}y^{\prime}t +
\frac{1}{4}(x^{\prime2}\!+\!y^{\prime2}) t^2 \right]e^{-t},
\label{rate}
\end{equation}
where $x^{\prime}$ and $y^{\prime}$ are phase rotations of $x$ and $y$ given by
$x^{\prime}=x\cos\delta+y\sin\delta$ and $y^{\prime}=y\cos\delta-x\sin\delta$
with $\delta$ the strong force phase between the CF and DCS processes.  
Clearly, Eq.~\eqref{rate} indicates that in the case of significant mixing the 
measured branching ratio is dependent on the lifetime acceptance of the 
analysis.  To account for this effect a large Monte Carlo sample of \rs decays 
is used.  Each Monte Carlo event accepted in the analysis is given a weight 
determined by the relative probability for an event with its lifetime given by 
Eq.~\eqref{rate} divided by the probability for the same lifetime in the 
exponential decay rate used to generate the Monte Carlo.  In this way we derive 
a relationship for $\rdcs$ as a function of $x^{\prime}$ and $y^{\prime}$ which
depends only on the measured branching ratio, and the first and second moments 
of the accepted lifetime distribution in the Monte Carlo.  The functional 
dependence on $x^{\prime}$ in the experimentally allowed region is small, while 
the dependence on $y^{\prime}$ (shown in Fig.~\ref{mixing}a for the case of 
$x^{\prime}=0$) is large.  For comparison, the $y^{\prime}$ and $\rdcs$ ranges 
from CLEO II.V~\cite{Godang:1999yd} and the $y$ limit from FOCUS (discussed in 
Sect.~\ref{dgamma}) are also shown in this figure.

To determine limits on $x^{\prime}$, $y^{\prime}$ and $\rdcs$, the $D\rarr K\pi$
data is split into high and low lifetime samples.  The \ws branching ratio
determined in each sample can be used to generate high and low lifetime curves
in $y^{\prime}\,\rdcs$ space like the one shown in Fig.~\ref{mixing}a.  The
point where these curves cross indicates the preferred values of $y^{\prime}$ and
$\rdcs$.  To quantitatively determine the 95\% confidence level (CL) allowed 
range, we integrate the likelihood for all points in the space and assign the 
95\% CL boundary to the high and low values beyond which the total integrated 
likelihood is equal to 2.5\%.  In determining $y^{\prime}$ ($\rdcs$) limits the 
value of $x^{\prime}$ is set to zero (the value of $x^{\prime}$ with greatest 
likelihood) and we integrate over the entire allowed range $\rdcs$
($y^{\prime}$) variable.  Using this procedure we find preliminary limits of
\begin{eqnarray*}
& -0.124<y^{\prime}<-0.006 & \mbox{and}\\ 
& 0.43\%<\rdcs <1.73\%.  
\end{eqnarray*}
The large upper limit on $\rdcs$ and large negative lower limit on $y^{\prime}$ 
are the result of a second crossing point of the high and low lifetime curves.  
This property is an unfortunate side effect of using only one lifetime split -- 
the fitting procedure~\cite{Link:2001kr} and limited statistics prevent more
data splits.  Nevertheless, the second crossing and its associated limits are 
far outside the allowed region of CLEO~II.V~\cite{Godang:1999yd} and are also 
expected to be ruled out by the improved limits on $x^{\prime}$ and 
$y^{\prime}$ from FOCUS semileptonic mixing studies.

To determine a limit on $|x^{\prime}|$ we integrate over the entire allowed 
range of $y^{\prime}$ and $\rdcs$ obtaining the preliminary limit of
\begin{equation*}
|x^{\prime}|<0.039.  
\end{equation*}
We also determined a 95\% CL boundary in $x^{\prime}\,y^{\prime}$ space by 
integrating over all allowed values of $\rdcs$ and selecting the boundary line 
that is isometric in likelihood and contains 95\% of the total likelihood.  
This boundary is shown in Fig.~\ref{mixing}b.  Also shown in Fig~\ref{mixing}b
are the best existing limits from the semileptonic mixing 
(E791\cite{Aitala:1996vz}), lifetime differences (FOCUS) and \ws (CLEO~II.V).

\section{Conclusions}
While the recent measurements in \mixing mixing do not rise to the level of a
discovery, they do warrant further study.  A discovery of a non-zero $y$ at the 
percent level ($R_{\mix}\sim 10^{-4}$) would not necessarily indicate new
physics, but it would be unexpectedly large, and at the very least would lead 
to a deeper understanding of the processes involved in meson mediated mixing.  

The current data from FOCUS and CLEO~II.V suggest two possible scenarios.  The
first scenario is that there is $y$-like mixing at the few percent level and a
large ($\sim\pi$ radians) strong phase.  The second scenario is that some or
all of the recent measurements are fluctuations.  New measurements are required
to determine which of these two possibilities is correct.

\end{document}